\documentclass{iopart}
\usepackage{indentfirst}
\usepackage{amsfonts}
\usepackage{amssymb}
\usepackage{bm}
\usepackage{a4wide}
\usepackage{graphicx}
\usepackage{cite}
\usepackage{makeidx}
\usepackage{multicol}
\usepackage{iopams}
\usepackage{bm}
\usepackage{graphicx}  % got figures? uncomment this
\begin{document}
\title[... neutrino energy quantization in rotating media]{Neutrino magnetic moment and neutrino energy quantization in rotating
media}
\author{I.~Balantsev}
\address{Department of Theoretical Physics, Moscow
State University, 119992 Moscow, Russia} \ead{laktan86@mail.ru}
\author{Yu.~Popov}
\address{Skobeltsyn Institute of Nuclear
Physics, Moscow State University, 119992 Moscow, Russia}
\ead{popov@srd.sinp.msu.ru}
\author{A.~Studenikin}
\address{Department of Theoretical Physics,
Moscow State University, 119992 Moscow, Russia}
\ead{studenik@srd.sinp.msu.ru}

%\maketitle
\begin{abstract}
After a brief discussion on neutrino electromagnetic properties, we
consider the problem of neutrino energy spectra in different media.
It is shown that in two particular cases (i.e., neutrino propagation
in a) transversally moving with increasing speed medium and b)
rotating medium) neutrino energies are quantized. These phenomena can
be important for astrophysical applications, for instance, for
physics of rotating neutron stars.
\end{abstract}
\vspace{2pc}
\section{Introduction - neutrino electromagnetic properties}
Initially the problem considered in this paper has originated from
the studies of neutrino electromagnetic properties and related items.
There is no doubt that the recent experimental and theoretical
studies of flavour conversion in solar, atmospheric, reactor and
accelerator neutrino fluxes give strong evidence of non-zero neutrino
mass.  A massive neutrino can have non-trivial electromagnetic
properties \cite{MarSanPLB77LeeShrPRD77FujShrPRL80}. A recent review
on neutrino electromagnetic properties can be found in
\cite{GiuStuPAN09}. The present situation in the domain is
characterized by the fact that in spite of reasonable efforts in
studies of a neutrino electromagnetic properties, there is no any
experimental confirmation in favour of a neutrino electromagnetic
characteristics being nonvanish. However, it is very plausible to
assume that a neutrino may have nonzero electromagnetic properties.
In particular, it seem very reasonable that neutrino have
nonvanishing magnetic moment \cite{GiuStuPAN09, StuNPB08}.

{\it Neutrino magnetic moment  interaction effects.} If a neutrino has
non-trivial electromagnetic properties, notably non-vanishing magnetic (and
also electric (transition) dipole moments or non-zero millicharge and charge
radius), then a direct neutrino couplings to photos become possible and several
important for applications processes exist \cite{RafPRL90}. A set of typical
and most important neutrino electromagnetic processes involving the direct
neutrino couplings with photons is: 1) a neutrino radiative decay
$\nu_{1}\rightarrow \nu_{2} +\gamma$, neutrino Cherenkov radiation in external
environment (plasma and/or electromagnetic fields), 2) photon (plasmon) decay
to a neutrino-antineutrino pair in plasma $\gamma \rightarrow \nu {\bar \nu }$,
\ 3) neutrino scattering off electrons (or nuclei), 4) neutrino spin
(spin-flavor) precession in magnetic field. Another very important phenomena is
the resonant amplification of the neutrino spin-flavour oscillations in matter
that was first considered in \cite{LimMarPRD88_AkhPLB88}.

Note a new mechanism of electromagnetic radiation produced by a neutrino moving
in matter and originated due to neutrino magnetic moment
\cite{LobStuPLB03LobStuPLB04DvoGriStuIJMP05,
StuTerPLB05GriStuTerPLB05GriStuTerG&C05GriStuTerPAN06,LobPLB05LobDAN05}.
It was termed the spin light of neutrino in matter ($SL\nu$)
\cite{LobStuPLB03LobStuPLB04DvoGriStuIJMP05}.  Although the $SL\nu$ was
considered first within quasiclassical approach, it was clear that this is a
quantum phenomenon by its nature. The quantum theory of this radiation has been
elaborated \cite{StuTerPLB05GriStuTerPLB05GriStuTerG&C05GriStuTerPAN06} (see
also \cite{LobPLB05LobDAN05}) within development
\cite{StuJPA08,StuJPA_06GriStuTerPAN09} of a quite powerful method that implies
the use of the exact solutions of the modified Dirac equation for the neutrino
wave function in matter. For elaboration of the quantum theory of the $SL\nu$
one has to find the solution of the quantum equation for the neutrino wave
function and for the neutrino energy spectrum in medium.

\section{Quantum equation for neutrino in medium}

The modified Dirac equation for the neutrino wave function exactly
accounting for the neutrino interaction with matter
\cite{StuTerPLB05GriStuTerPLB05GriStuTerG&C05GriStuTerPAN06}:
\begin{equation}\label{new} \Big\{
i\gamma_{\mu}\partial^{\mu}-\frac{1}{2}
\gamma_{\mu}(1+\gamma_{5})f^{\mu}-m \Big\}\Psi(x)=0.
\end{equation}
This is the most general form of the equation for the neutrino wave
function in which the effective potential
$V_{\mu}=\frac{1}{2}(1+\gamma_{5})f_{\mu}$ includes both the neutral
and charged current interactions of neutrino with the background
particles and which can also account for effects  of matter motion
and polarization. It should be mentioned that other modifications of
the Dirac equation were previously used in \cite{Man88_Zha91} for
studies of the neutrino dispersion relations, neutrino mass
generation and neutrino oscillations in the presence of matter.

In the case of matter composed of electrons, neutrons, and protons
and for the neutrino interaction with background particles given by
the standard model supplied with the singlet right-handed neutrino
one has
\begin{equation}\label{Lag_f}
f^\mu={\sqrt2 G_F }\sum\limits_{f=e,p,n}
j^{\mu}_{f}q^{(1)}_{f}+\lambda^{\mu}_{f}q^{(2)}_{f},
\end{equation}
where
\begin{equation}\label{q_f}
 q^{(1)}_{f}= (I_{3L}^{(f)}-2Q^{(f)}\sin^{2}\theta_{W}+\delta_{ef}),
\ q^{(2)}_{f}=-(I_{3L}^{(f)}+\delta_{ef}), \  \delta_{ef}=\left\{
\begin{tabular}{rcl}
1 \ \ \ for {\it f=e}, \\
0 \ \ \ for {\it f=n, p}. \\
\end{tabular}
\right.
\end{equation}
Here $I_{3L}^{(f)}$ and $Q^{(f)}$ are, respectively,  values of the
isospin third components and the electric charges of matter particles
($f=e,n,p$). The corresponding currents $j_{f}^{\mu}$ and
polarization vectors $\lambda_{f}^{\mu}$ are
\begin{equation}\label{j}
j_{f}^\mu=(n_f,n_f{\bf v}_f),
%\begin{equation} \label{lambda}
\ \ \ \lambda_f^{\mu} =\Bigg(n_f ({\bm \zeta}_f {\bf v}_f ), n_f {\bm
\zeta}_f \sqrt{1-v_f^2}+ {{n_f {\bf v}_f ({\bm \zeta}_f {\bf v}_f )}
\over {1+\sqrt{1- v_f^2}}}\Bigg),
\end{equation}
where $\theta _{W}$ is the Weinberg angle. In the above formulas
(\ref{j}), $n_f$, ${\bf v}_f$ and ${\bm \zeta}_f \ (0\leq |{\bm
\zeta}_f |^2 \leq 1)$ stand, respectively, for the invariant number
densities, average speeds and polarization vectors of the matter
components.

In the case of matter at rest it is possible to solve the modified
Dirac equation for different types of neutrinos moving in matter of
different composition, as it is shown in
\cite{StuTerPLB05GriStuTerPLB05GriStuTerG&C05GriStuTerPAN06}. The
energy spectrum of different neutrinos moving in matter  is given by
\begin{equation}\label{Energy}
  E_{\varepsilon}=\varepsilon \eta {\sqrt{{\bf p}^{2}\Big(1-s\alpha \frac{m}{p}\Big)^{2}
  +m^2} +\alpha m}.
\end{equation}
In the general case of matter composed of electrons, neutrons and
protons the matter density parameter $\alpha$ for different neutrino
species is
\begin{equation}\label{alpha}
  \alpha_{\nu_e,\nu_\mu,\nu_\tau}=
  \frac{1}{2\sqrt{2}}\frac{G_F}{m}\Big(n_e(4\sin^2 \theta
_W+\varrho)+n_p(1-4\sin^2 \theta _W)-n_n\Big),
\end{equation}
where $\varrho=1$ for the electron neutrino and $\varrho=-1$ for the
muon and tau neutrinos.

The value $\eta=$sign$\big(1-s\alpha\frac{m}{p}\big)$ in (\ref{Energy})
provides a proper behavior of the wave function in the hypothetical massless
case. The values $s=\pm 1$ specify the two neutrino helicity states, $\nu_{+}$
and  $\nu_{-}$. The quantity $\varepsilon=\pm 1$ splits the solutions into the
two branches that in the limit of the vanishing matter density,
$\alpha\rightarrow 0$, reproduce the positive- and negative-frequency
solutions, respectively.

In the next two sections we apply the developed method of exact solutions to
two particular cases when neutrino is propagating in transversally moving with
increasing speed medium \cite{GriSavStuIzvVuz07, StuJPA08} and a rotating
medium of constant density. In both cases the obtained energy spectrum of
neutrino is quantized like the energy spectrum of an electron is quantized in a
constant magnetic field.
\section{Neutrino quantum states in transversally moving with
increasing speed medium}

First we consider a neutrino propagating in medium composed of
neutrons that move perpendicular to the neutrino path with linearly
increasing speed. This can be regarded as the first approach to
modelling of neutrino propagation inside a rotating neutron star
\cite{GriSavStuIzvVuz07, StuJPA08}. The corresponding modified Dirac
equation for the neutrino wave function is given by (\ref{new}) with
the matter potential accounting for rotation,
\begin{equation}\label{rot_f}
f^\mu = -{G}(n,n {\bf v}), \ \ {\bf v}=(\omega y,0,0),
\end{equation}
where $G=\frac{G_F}{\sqrt{2}}$. Here $\omega$ is the angular
frequency of matter rotation around OZ axis, it also is supposed that
the neutrino propagates along OY axis. For  the neutrino wave
function components $\psi (x) $ we get from the modified Dirac
equation (\ref{new}), a set of equations{\footnote{The chiral
representation for Dirac matrixes is used.},
\begin{equation}
\label{Dir_Comp}
%\left\{
\begin{array}
{rcl} \big[i\left(\partial_0 - \partial_3\right) + G n\big] \psi_1 +
\big[-\left(i\partial_1 + \partial_2\right) + G n \omega y \big]
\psi_2
= m \psi_3, \\
\big[\left(-i\partial_1 + \partial_2 \right) + G n \omega y \big]
\psi_1 + \big[i\left(\partial_0 +
\partial_3\right) + G n \big] \psi_2 = m \psi_4,\\
i\left( \partial_0 + \partial_3 \right) \psi_3 + \left(i
\partial_1 +
\partial_2 \right) \psi_4 = m \psi_1,\\
\left(i \partial_1 - \partial_2\right) \psi_3 + i \left(\partial_0 -
\partial_3 \right) \psi_4 = m \psi_2.\end{array}
%\right.
\end{equation}
In general case, it is not a trivial task to find solutions of this
set of equations.

The problem is reasonably simplified in the limit of a very small
neutrino mass, i.e. when the neutrino mass can be ignored in the
left-hand side of (\ref{Dir_Comp})  in respect to the kinetic and
interaction terms in the right-hand sides of these equations. In this
case two pairs of the neutrino wave function components decouple one
from each other and four equations (\ref{Dir_Comp}) split into the
two independent sets of two equations, that couple together the
neutrino wave function components in pairs, ($\psi_1,\ \psi_2$) and
($\psi_3,\ \psi_4$).

The second pair of equations (\ref{Dir_Comp}) does not contain a
matter term and is attributed to the sterile right-handed chiral
neutrino state, $\psi_{R}$. The corresponding solution can be taken
in the plain-wave form
\begin{equation}
\psi_{R} \sim L^{-\frac{3}{2}}\exp \{i(- p_0 t + p_1 x + p_2 y + p_3
z)\}\psi,
\end{equation}
where $p_\mu$ is the neutrino momentum. Then for the components
$\psi_3$ and $\psi_4$ we obtain from (\ref{Dir_Comp}) the following
equations
\begin{equation}
\label{PsiR_comp}
%\left\{
\begin{array}{rcl}
\left( p_0 - p_3 \right) \psi_3 - \left( p_1 - i p_2\right) \psi_4 = 0, \\

-\left( p_1 + i p_2 \right) \psi_3 + \left( p_0 + p_3\right) \psi_4 =
0.
\end{array}
%\right.
\end{equation}
Finally, from (\ref{PsiR_comp}) for the sterile right-handed neutrino
we get
\begin{equation}
\label{PsiR} \psi_R = \frac{\mathrm{e}^{-i p
x}}{L^{3/2}\sqrt{2p_0(p_0 - p_3)}}\left(
\begin{array}{c}
0 \\
0 \\
- p_1 + i p_2 \\
p_3 - p_0
\end{array}
\right),
\end{equation}
where $px=p_{\mu}x^{\mu}, \ p_{\mu}=(p_0,p_1,p_2,p_3)$ and $x_{\mu} =
(t,x,y,z)$. This solution, as it should do, has the vacuum dispersion
relation.

In the neutrino mass vanishing limit the first pair of equations
(\ref{Dir_Comp}) corresponds to the active left-handed neutrino. The
form of these equations is similar to the correspondent equations for
a charged particle (e.g., an electron) moving in a constant magnetic
field $B$ given by the potential $\bm{A} = (By,0,0)$ (see, for
instance, \cite{SokTerSynRad68}). To display
 the analogy, we note that in our case the matter
current component $n\bm{v}$ plays the role of the vector potential
$\bm{A}$. The existed analogy between an electron dynamics in an
external electromagnetic field and a neutrino dynamics in background
matter is discussed in \cite{StuJPA08}.

The  solution of the first pair of equations (\ref{Dir_Comp}) can be
taken in the form
\begin{equation}
\psi_{L} \sim \frac{1}{L}\exp \{i(- p_0 t + p_1 x + p_3 z)\}\psi(y),
\end{equation}
and for the components $\psi_1$ and $\psi_2$ of the neutrino wave
function we obtain from (\ref{Dir_Comp}) the following equations
\begin{equation}
\label{PsiL_comp}
\begin{array}{rcl}
\Big( p_0 + p_3 + G n \Big) \psi_1 - \sqrt{\rho}
\left(\frac{\partial }{\partial \eta} - \eta\right) \psi_2 = 0, \\
\sqrt{\rho}\left(\frac{\partial }{\partial \eta} + \eta\right) \psi_1
+ \Big(p_0 - p_3 + G n \Big) \psi_2 = 0,
\end{array}
\end{equation}
where
\begin{eqnarray}
\label{eta} \eta =  \sqrt{\rho} \left(x_2 + \frac{ p_1}{\rho} \right)
, \ \  \rho= G n \omega.
\end{eqnarray}
For the wave function we finally get
\begin{equation} \label{PsiL}
\psi_L = \frac{\rho^{\frac{1}{4}}\mathrm{e}^{-ip_0 t + ip_1x + ip_3
z}}{L\sqrt{(p_0 - p_3 + G n)^2+2\rho N}}\left(
\begin{array}{c}
\left(p_0 - p_3 + G n\right) u_N(\eta) \\
- \sqrt{2\rho N} u_{N-1}(\eta) \\
0 \\
0 \end{array} \right),
\end{equation}
where $u_N(\eta)$ are Hermite functions of order $N$. For the energy
of the active left-handed neutrino  we get
\begin{equation}\label{nu_quant_energy}
\label{energy_L} p_0 = \sqrt{p_3^2 + 2 \rho N} - G n, \ \ N=0,1,2,...
\ .
\end{equation}
The energy depends on the neutrino momentum component $p_3$ along the
rotation axis of matter and the quantum number $N$ that determines
the magnitude of the neutrino momentum in the orthogonal plane. For
description of antineutrinos one has to consider the ``negative
sign``  energy eigeinvalues (see similar discussion in Section 2.2).
Thus, the energy of an electron antineutrino in the rotating matter
composed of neutrons is given by
\begin{equation}\label{antinu_quant_energy}
\label{energy_L} \tilde p_0 = \sqrt{p_3^2 + 2 \rho N} + G n, \ \
N=0,1,2,... \ .
\end{equation}
Obviously, generalization for different other neutrino flavours and
matter composition is just straightforward (see (\ref{j}) and
(\ref{alpha})).

Thus, it is shown \cite{GriSavStuIzvVuz07} that the transversal
motion of an active neutrino and antineutrino is quantized in moving
matter very much like an electron energy is quantized in a constant
magnetic field that corresponds to the relativistic form of the
Landau energy levels (see \cite{SokTerSynRad68}).

\section{Neutrino energy in a rotating medium}

Now we consider a more consistent model of a neutrino motion in a rotating
matter. For this case we choose the effective matter potential in (\ref{new})
in the following form
\begin{equation}\label{real_rot_f}
f^\mu = -{G}(n,n {\bf v}), \ \ {\bf v}=(-\omega y,\omega x,0).
\end{equation}
Contrary to the case considered in the previous section, equation
(\ref{new}) with the potential (\ref{real_rot_f}) describes the case
when a neutrino is moving in a rotating media. It is shown below how
the equation (\ref{new}) with (\ref{real_rot_f}) can be solved and
the corresponding neutrino energy spectrum is obtained.

The solution of the equation (\ref{new}) with (\ref{real_rot_f}) can
be sought in the form
\begin{equation}\label{psi}
\psi(t,x,y,z) = \mathrm{e}^{-ip_0 t + ip_3z}\left(
\begin{array}{c}
\psi_1(x,y)\\
\psi_2(x,y)\\
\psi_3(x,y)\\
\psi_4(x,y)\end{array} \right).
\end{equation}
Substituting (\ref{psi}) into  (\ref{new}) with (\ref{real_rot_f})
and using the explicit form of the Dirac matrices in the chiral
representation, we arrive at a system of linear equations for the
neutrino wave function components:

\begin{equation}\label{system}
\begin{array}{rcl}
-(p_0+p_3+Gn)\psi_1+i\left\{\left(\frac{\partial}{\partial x}-
i\frac{\partial}{\partial y}\right)+Gn\omega(x-iy)\right\}\psi_2=-m\psi_3,\\
i\left\{\left(\frac{\partial}{\partial x}+i\frac{\partial}{\partial y}\right)-
Gn\omega(x+iy)\right\}\psi_1+(p_3-p_0-Gn)\psi_2=-m\psi_4,\\
(p_0-p_3)\psi_3+i\left(\frac{\partial}{\partial x}-i\frac{\partial}
{\partial y}\right)\psi_4=m\psi_1,\\
+i\left(\frac{\partial}{\partial x}+i\frac{\partial}{\partial
y}\right)\psi_3+(p_0+p_3)\psi_4=m\psi_2.
\end{array}
\end{equation}

In the polar coordinates $x+iy=r \mathrm{e}^{i\phi},  x-iy=r
\mathrm{e}^{-i\phi}$ one has
\begin{equation}
\begin{array}{rcl}
\frac{\partial}{\partial x}+i\frac{\partial}{\partial
y}=\mathrm{e}^{i\phi}\left(\frac{\partial}{\partial r}+\frac{i}{r}
\frac{\partial}{\partial\phi}\right), \ \ \ \frac{\partial}{\partial
x}-i\frac{\partial}{\partial y}=
\mathrm{e}^{-i\phi}\left(\frac{\partial}{\partial r}-\frac{i}{r}
\frac{\partial}{\partial\phi}\right),
\end{array}
\end{equation}
and the system of equations (\ref{system}) transforms to
\begin{equation}\label{system_1}
\begin{array}{rcl}
-(p_0+p_3+Gn)\psi_1+i\mathrm{e}^{-i\phi}\left\{\frac{\partial}
{\partial r}-\frac{i}{r}\frac{\partial}{\partial\phi}+\rho r\right\}\psi_2=-m\psi_3,\\
i\mathrm{e}^{i\phi}\left\{\frac{\partial}{\partial r}+\frac{i}
{r}\frac{\partial}{\partial\phi}-\rho r\right\}\psi_1+(p_3-p_0-Gn)\psi_2=-m\psi_4,\\
(p_0-p_3)\psi_3+i\mathrm{e}^{-i\phi}\left(\frac{\partial}{\partial
r}-
\frac{i}{r}\frac{\partial}{\partial\phi}\right)\psi_4=m\psi_1,\\
+i\mathrm{e}^{i\phi}\left(\frac{\partial}{\partial
r}+\frac{i}{r}\frac{\partial}{\partial\phi}\right)\psi_3+(p_0+p_3)\psi_4=m\psi_2.
\end{array}
\end{equation}

It is possible to show that the operator of the total momentum $J_z=L_z+S_z$,
where $L_z=-i\frac{\partial}{\partial\phi}$, $S_z=\frac{1}{2}\sigma_3$,
commutes with the corresponding Hamiltonian of the considered system. Therefore
the solutions can be taken in the form
\begin{equation}\left(
\begin{array}{c}
\psi_1\\\psi_2\\\psi_3\\\psi_4\end{array}\right)
=\left(\begin{array}{c}i\chi_1(r)e^{i(l-1)\phi}\\\chi_2(r)e^{il\phi}\\
i\chi_3(r)e^{i(l-1)\phi}\\\chi_4(r)e^{il\phi}\end{array}\right),\label{form
of solution}
\end{equation}
that are the eigenvectors for the total momentum operator $J_z$ with
the corresponding eigenvalues $l-\frac{1}{2}$. After substitution of
(\ref{form of solution}) the system (\ref{system_1}) can be rewritten
in the following form,
\begin{equation}\label{system_2}
\begin{array}{rcl}
-(p_0+p_3+Gn)\chi_1+\left\{\frac{d}{d r}+\frac{l}{r}+\rho r\right\}\chi_2=-m\chi_3,\\
\left\{\frac{d}{d r}-\frac{l-1}{r}-\rho r\right\}\chi_1+(p_0-p_3+Gn)\chi_2=m\chi_4,\\
(p_0-p_3)\chi_3+\left(\frac{d}{d r}+\frac{l}{r}\right)\chi_4=m\chi_1,\\
\left(\frac{d}{d r}-\frac{l-1}{r}\right)\chi_3-(p_0+p_3)\chi_4=-m\chi_2.
\end{array}
\end{equation}
For the further consideration it is convenient to introduce the
 rising and decreasing operators
\begin{eqnarray}
R^+=\frac{d}{d r}-\frac{l-1}{r}-\rho r, \ \ \ \
R^-=\frac{d}{d r}+\frac{l}{r}+\rho r.
\end{eqnarray}
After applications of the decreasing $R^-$ and increasing $R^+$
operators to the second and first equations of (\ref{system_1})
correspondently, one gets
\begin{equation}\label{system_3}
\begin{array}{rcl}
R^-R^+\chi_1+\left((p_0+Gn)^2-p_3^2-m^2\right)\chi_1=m(Gn\chi_3+\rho r\chi_4),\\
R^+R^-\chi_2+\left((p_0+Gn)^2-p_3^2-m^2\right)\chi_2=m(Gn\chi_4+\rho r\chi_3).
\end{array}
\end{equation}

Note that the system (\ref{system_2}), as well as the system (\ref{Dir_Comp}),
can be solved exactly in the limit of vanishing neutrino mass $m\rightarrow 0$.
In order to find the non-zero-mass correction to the energy spectrum of a
neutrino in a bound state in matter, the neutrino square integrable
wavefunction should be found. Therefore, in analogy with the zero-mass case we
take $\chi_3=\chi_4=0$ in the lowest order of perturbation series expansion.
Thus we arrive to the system
\begin{equation}\label{reduced equations}
\begin{array}{rcl}
R^-R^+\chi_1+\left((p_0+Gn)^2-p_3^2-m^2\right)\chi_1=0,\\
R^+R^-\chi_2+\left((p_0+Gn)^2-p_3^2-m^2\right)\chi_2=0.
\end{array}
\end{equation}
The solution of (\ref{reduced equations}) can be written in the form
\begin{equation}\label{solution}
\left(
\begin{array}{c}
\chi_1\\ \chi_2\end{array}\right)=\left(
\begin{array}{c}
C_1\mathcal{L}_s^{l-1}(\rho r^2)\\C_2\mathcal{L}_s^l(\rho
r^2)\end{array}\right),
\end{equation}
where $\mathcal{L}_s^l$ are the Laguerre functions
\cite{SokTerSynRad68}. After substitution of (\ref{solution}) to
(\ref{reduced equations}) and taking into account the properties of
the increasing and decreasing rising operators,
\begin{equation}
\begin{array}{rcl}
R^+\,\mathcal{L}_s^{l-1}(\rho r^2)=-2\sqrt{\rho(s+l)}\,\mathcal{L}_s^l(\rho
r^2), \\  R^-\,\mathcal{L}_s^{l}(\rho
r^2)=2\sqrt{\rho(s+l)}\,\mathcal{L}_s^{l-1}(\rho r^2),
\end{array}
\end{equation}
we get from (\ref{reduced equations}) the  equation for the neutrino
energy spectrum in matter
\begin{equation}\label{ne_energy_spectrum_matter}
m^2+p_3^2+4(s+l)\rho-(p_0+Gn)^2=0.
\end{equation}
Solving this equation we get for the neutrino energies
\begin{equation}
p_0=\pm\sqrt{m^2+p_3^2+4N\rho}-Gn, \ \ N=0,1,2,... \ .\label{energy
spectrum}
\end{equation}
where the quantum number $N=s+l$ is introduced. As usually, two signs
in the solution corresponds to the neutrino and antineutrino
energies, correspondently,
\begin{equation}\label{new_energy}
p_0=\sqrt{m^2+p_3^2+4N\rho}-Gn, \ \ {\tilde
p}_0=\sqrt{m^2+p_3^2+4N\rho}+Gn, .\label{energy spectrum}
\end{equation}

From the obtained energy spectrum it just straightforward that the transversal
motion momentum of an antineutrino is given by
\begin{equation}
\tilde p_{\bot}= 2\sqrt{NG\omega}.
\end{equation}
The quantum number $N$ determines also the radius of the quasiclassical orbit
in matter (it is supposed that $N\gg 1$ and $p_3=0$),
\begin{equation}
R=\sqrt{\frac{N}{G n \omega}}.
\end{equation}
It follows that antineutrinos can have bound orbits inside a rotating
star. To make an estimation of magnitudes, let us consider a model of
a rotating neutron star with radius $R_{NS}=10 \ km$, matter density
$n=10^{37} cm^{-3}$ and angular frequency $\omega=2\pi \times 10^{3}
\ s^{-1}$. For this set of parameters, the radius of an orbit is less
than the typical star radius $R_{NS}$ if the quantum number $N\leq
N_{max} =10^{10}$. Therefore, antineutrinos that occupy orbits with
$N\leq 10^{10}$ can be bounded inside the star. The scale of the
bounded antineutrinos energy estimated by (\ref{new_energy}) is of
the order $\tilde p_0\sim 1 \ eV$. It should be underlined that
within the quasiclassical approach the neutrino binding on circular
orbits is due to an effective force that is orthogonal to the
particle speed. Note that there is another mechanism of neutrinos
binding inside a neutron star when the effect is produced by a
gradient of the matter density (see the last paper in
\cite{Man88_Zha91}). A discussion on the ``matter-induced Lorentz
force`` that can be introduced in order to explain a neutrino motion
on quasiclassical circular orbits can be found in \cite{StuJPA08}.

We ague that the effect of a neutrino energy quantization can have important
consequences for physics of rotating neutron stars.

{\it Acknowledgements.} One of the authors (A.S.) thanks Giorgio
Bellettini, Giorgio Chiarelli, Mario Greco and Gino Isidori for the
kind invitation to participate in the XXIII Recontres de Physique de
La Vallee D'Aoste on Results and Perspectives in Particle Physics and
also thanks all the organizers for their hospitality in La Thuile.
The authors are also thankful to Alexander Grigoriev for fruitful
discussions.

\end{document}